\documentclass[aps,prc,twocolumn,superscriptaddress,altaffilletter]{revtex4-1}

\usepackage{graphicx}
\usepackage{epstopdf}
\usepackage{multirow}
\usepackage[pagebackref=false]{hyperref}	
\hypersetup{
  colorlinks = true,
  linkcolor=blue,   
  citecolor=blue,   
  urlcolor=blue,    
  pdfdisplaydoctitle=true
}

\newcommand{\gosia}{\textsc{gosia}}	

\begin{document}

\pagebreak
\title{Collectivity in the light radon nuclei measured directly via Coulomb excitation}

\author{L.~P.~Gaffney}
\email[Corresponding author: ]{Liam.Gaffney@fys.kuleuven.be}
\affiliation{KU Leuven, Instituut voor Kern- en Stralingsfysica, 3001 Leuven, Belgium}
\affiliation{Oliver Lodge Laboratory, University of Liverpool, Liverpool L69 7ZE, United Kingdom}

\author{A.~P.~Robinson}
\affiliation{Department of Physics, The University of York, Heslington, York YO10 5DD, United Kingdom}
\affiliation{School of Physics and Astronomy, The University of Manchester, Manchester M13 9PL, United Kingdom}

\author{D.~G.~Jenkins}
\affiliation{Department of Physics, The University of York, Heslington, York YO10 5DD, United Kingdom}

\author{A.~N.~Andreyev}
\affiliation{KU Leuven, Instituut voor Kern- en Stralingsfysica, 3001 Leuven, Belgium}
\affiliation{Department of Physics, The University of York, Heslington, York YO10 5DD, United Kingdom} %
\affiliation{Advanced Science Research Centre, Japan Atomic Energy Agency, Tokai-mura, 319-1195, Japan} %

\author{M.~Bender}
\affiliation{Universit\'{e} Bordeaux, Centre d'Etudes Nucl\'{e}aires de Bordeaux Gradignan, UMR5797, F-33175 Gradignan, France}
\affiliation{CNRS/IN2P3, Centre d'Etudes Nucl\'{e}aires de Bordeaux Gradignan, UMR5797, F-33175 Gradignan, France}

\author{A.~Blazhev}
\affiliation{Institut f\"ur Kernphysik, Universit\"at zu K\"oln, 50937 K\"oln, Germany}

\author{N.~Bree}
\affiliation{KU Leuven, Instituut voor Kern- en Stralingsfysica, 3001 Leuven, Belgium}

\author{B.~Bruyneel}
\affiliation{Institut f\"ur Kernphysik, Universit\"at zu K\"oln, 50937 K\"oln, Germany}

\author{P.~A.~Butler}
\affiliation{Oliver Lodge Laboratory, University of Liverpool, Liverpool L69 7ZE, United Kingdom}

\author{T.~E.~Cocolios}
\affiliation{ISOLDE, CERN, CH-1211 Geneva 23, Switzerland}
\affiliation{School of Physics and Astronomy, The University of Manchester, Manchester M13 9PL, United Kingdom} %

\author{T.~Davinson}
\affiliation{Department of Physics and Astronomy, University of Edinburgh, EH9 3JZ, United Kingdom}

\author{A.~N.~Deacon}
\affiliation{School of Physics and Astronomy, The University of Manchester, Manchester M13 9PL, United Kingdom}

\author{H.~De~Witte}
\affiliation{KU Leuven, Instituut voor Kern- en Stralingsfysica, 3001 Leuven, Belgium}

\author{D.~DiJulio}
\affiliation{Physics Department, University of Lund, Box 118, SE-221 00 Lund, Sweden}

\author{J.~Diriken}
\affiliation{KU Leuven, Instituut voor Kern- en Stralingsfysica, 3001 Leuven, Belgium}

\author{A.~Ekstr\"{o}m}
\affiliation{Physics Department, University of Lund, Box 118, SE-221 00 Lund, Sweden}

\author{Ch.~Fransen}
\affiliation{Institut f\"ur Kernphysik, Universit\"at zu K\"oln, 50937 K\"oln, Germany}

\author{S.~J.~Freeman}
\affiliation{School of Physics and Astronomy, The University of Manchester, Manchester M13 9PL, United Kingdom}

\author{K.~Geibel}
\affiliation{Institut f\"ur Kernphysik, Universit\"at zu K\"oln, 50937 K\"oln, Germany}

\author{T.~Grahn}
\affiliation{University of Jyvaskyla, Department of Physics, P.O. Box 35, FI-40014 University of Jyvaskyla, Finland}
\affiliation{Helsinki Institute of Physics, University of Helsinki, P.O. Box 64, FIN-00014 Helsinki, Finland}

\author{B.~Hadinia}
\affiliation{School of Engineering, University of the West of Scotland, Paisley PA1 2BE, United Kingdom}

\author{M.~Hass}
\affiliation{The Weizmann Institute, 76100 Rehovot, Israel}

\author{P.-H.~Heenen}
\affiliation{Physique Nucl\'{e}aire Th\'{e}orique, Universit\'{e} Libre de Bruxelles, C.P. 229, B-1050 Bruxelles, Belgium}

\author{H.~Hess}
\affiliation{Institut f\"ur Kernphysik, Universit\"at zu K\"oln, 50937 K\"oln, Germany}

\author{M.~Huyse}
\affiliation{KU Leuven, Instituut voor Kern- en Stralingsfysica, 3001 Leuven, Belgium}

\author{U.~Jakobsson}
\altaffiliation[Present address: ]{KTH, The Division of Nuclear Physics, AlbaNova University center, SE-106 91 Stockholm, Sweden} %
\affiliation{University of Jyvaskyla, Department of Physics, P.O. Box 35, FI-40014 University of Jyvaskyla, Finland}
\affiliation{Helsinki Institute of Physics, University of Helsinki, P.O. Box 64, FIN-00014 Helsinki, Finland}

\author{N.~Kesteloot}
\affiliation{KU Leuven, Instituut voor Kern- en Stralingsfysica, 3001 Leuven, Belgium}
\affiliation{Belgian Nuclear Research Centre SCK$\bullet$CEN, B-2400 Mol, Belgium}

\author{J.~Konki}
\affiliation{ISOLDE, CERN, CH-1211 Geneva 23, Switzerland}
\affiliation{University of Jyvaskyla, Department of Physics, P.O. Box 35, FI-40014 University of Jyvaskyla, Finland}
\affiliation{Helsinki Institute of Physics, University of Helsinki, P.O. Box 64, FIN-00014 Helsinki, Finland}

\author{Th.~Kr\"{o}ll}
\affiliation{Institut f\"{u}r Kernphysik, Technische Universit\"{a}t Darmstadt, D-64289 Darmstadt, Germany}

\author{V.~Kumar}
\affiliation{The Weizmann Institute, 76100 Rehovot, Israel}

\author{O.~Ivanov}
\affiliation{KU Leuven, Instituut voor Kern- en Stralingsfysica, 3001 Leuven, Belgium}

\author{S.~Martin-Haugh}
\affiliation{Department of Physics, The University of York, Heslington, York YO10 5DD, United Kingdom}

\author{D.~M\"{u}cher}
\affiliation{Physik Department E12, Technische Universit\"{a}t M\"{u}nchen, D-85748 Garching, Germany}

\author{R.~Orlandi}
\affiliation{School of Engineering, University of the West of Scotland, Paisley PA1 2BE, United Kingdom}
\affiliation{Advanced Science Research Center, Japan Atomic Energy Agency, Tokai, Ibaraki, 319-1195, Japan} %

\author{J.~Pakarinen}
\affiliation{ISOLDE, CERN, CH-1211 Geneva 23, Switzerland}
\affiliation{University of Jyvaskyla, Department of Physics, P.O. Box 35, FI-40014 University of Jyvaskyla, Finland}
\affiliation{Helsinki Institute of Physics, University of Helsinki, P.O. Box 64, FIN-00014 Helsinki, Finland}

\author{A.~Petts}
\affiliation{Oliver Lodge Laboratory, University of Liverpool, Liverpool L69 7ZE, United Kingdom}

\author{P.~Peura}
\affiliation{University of Jyvaskyla, Department of Physics, P.O. Box 35, FI-40014 University of Jyvaskyla, Finland}
\affiliation{Helsinki Institute of Physics, University of Helsinki, P.O. Box 64, FIN-00014 Helsinki, Finland}

\author{P.~Rahkila}
\affiliation{University of Jyvaskyla, Department of Physics, P.O. Box 35, FI-40014 University of Jyvaskyla, Finland}
\affiliation{Helsinki Institute of Physics, University of Helsinki, P.O. Box 64, FIN-00014 Helsinki, Finland}

\author{P.~Reiter}
\affiliation{Institut f\"ur Kernphysik, Universit\"at zu K\"oln, 50937 K\"oln, Germany}

\author{M.~Scheck}
\affiliation{Oliver Lodge Laboratory, University of Liverpool, Liverpool L69 7ZE, United Kingdom}
\affiliation{School of Engineering, University of the West of Scotland, Paisley PA1 2BE, United Kingdom} %
\affiliation{Scottish Universities Physics Alliance, Glasgow G12 8QQ, United Kingdom} %

\author{M.~Seidlitz}
\affiliation{Institut f\"ur Kernphysik, Universit\"at zu K\"oln, 50937 K\"oln, Germany}

\author{K. Singh}
\affiliation{The Weizmann Institute, 76100 Rehovot, Israel}

\author{J.~F.~Smith}
\affiliation{School of Engineering, University of the West of Scotland, Paisley PA1 2BE, United Kingdom}

\author{J.~Van~de~Walle}
\affiliation{ISOLDE, CERN, CH-1211 Geneva 23, Switzerland}

\author{P.~Van~Duppen}
\affiliation{KU Leuven, Instituut voor Kern- en Stralingsfysica, 3001 Leuven, Belgium}

\author{D.~Voulot}
\affiliation{ISOLDE, CERN, CH-1211 Geneva 23, Switzerland}

\author{R.~Wadsworth}
\affiliation{Department of Physics, The University of York, Heslington, York YO10 5DD, United Kingdom}

\author{N.~Warr}
\affiliation{Institut f\"ur Kernphysik, Universit\"at zu K\"oln, 50937 K\"oln, Germany}

\author{F.~Wenander}
\affiliation{ISOLDE, CERN, CH-1211 Geneva 23, Switzerland}

\author{K.~Wimmer}
\altaffiliation[Present address: ]{Department of Physics, University of Tokyo, Hongo, Bunkyo-ku, Tokyo, 113-0033, Japan} %
\affiliation{Physik Department E12, Technische Universit\"{a}t M\"{u}nchen, D-85748 Garching, Germany}

\author{K.~Wrzosek-Lipska}
\affiliation{KU Leuven, Instituut voor Kern- en Stralingsfysica, 3001 Leuven, Belgium}
\affiliation{Heavy Ion Laboratory, University of Warsaw, PL-00-681 Warsaw, Poland} %

\author{M.~Zieli\'{n}ska}
\affiliation{Heavy Ion Laboratory, University of Warsaw, PL-00-681 Warsaw, Poland}
\affiliation{CEA Saclay, DAPNIA/SPhN, F-91191 Gif-sur-Yvette, France} %

\date{\today}

\begin{abstract}
\newlength{\abstractwidth}
\setlength{\abstractwidth}{\textwidth}	
\addtolength{\abstractwidth}{-4\leftmargin}
\addtolength{\abstractwidth}{2\parindent}
\setlength{\parindent}{-\parindent}
\begin{minipage}{\abstractwidth}
\begin{description}
\item[Background] Shape coexistence in heavy nuclei poses a strong challenge to state-of-the-art nuclear models, where several competing shape minima are found close to the ground state.
A classic region for investigating this phenomenon is in the region around $Z=82$ and the neutron mid-shell at $N=104$.
\item[Purpose] Evidence for shape coexistence has been inferred from $\alpha$-decay measurements, laser spectroscopy and in-beam measurements. While the latter allow the pattern of excited states and rotational band structures to be mapped out, a detailed understanding of shape coexistence can only come from measurements of electromagnetic matrix elements.
\item[Method] Secondary, radioactive ion beams of $^{202}$Rn and $^{204}$Rn were studied by means of low-energy Coulomb excitation at the REX-ISOLDE facility in CERN.
\item[Results] The electric-quadrupole ($E2$) matrix element connecting the ground state and first-excited $2^{+}_{1}$ state was extracted for both $^{202}$Rn and $^{204}$Rn, corresponding to ${B(E2;2^{+}_{1} \to 2^{+}_{1})=29^{+8}_{-8}}$~W.u. and~$43^{+17}_{-12}$~W.u., respectively.
Additionally, $E2$ matrix elements connecting the $2^{+}_{1}$ state with the $4^{+}_{1}$ and $2^{+}_{2}$ states were determined in $^{202}$Rn.
No excited $0^{+}$ states were observed in the current data set, possibly due to a limited population of second-order processes at the currently-available beam energies. 
\item[Conclusions] The results are discussed in terms of collectivity and the deformation of both nuclei studied is deduced to be weak, as expected from the low-lying level-energy schemes. Comparisons are also made to state-of-the-art beyond-mean-field model calculations and the magnitude of the transitional quadrupole moments are well reproduced.
\end{description}
\end{minipage}
\end{abstract}

\pacs{21.10.Ky, 21.21.Re, 23.20.Js, 25.70.De, 27.80.+w, 29.38.Gj}    
\maketitle

\section{Introduction}

Shape coexistence in nuclei is a phenomenon whereby two or more nucleon configurations, each with a different macroscopic shape, exist together at similar energy.
It has been observed in a number of regions of the nuclear chart and, over the last decade and more, extensive experimental evidence has been found in support of the shape coexistence in the Pb region~\cite{Heyde2011}.
The most-striking early indications came from isotope-shift measurements in mercury ($Z=80$), which showed a large discontinuity in the mean-square-charge radii between $^{185}$Hg and $^{187}$Hg~\cite{Bonn1972}.
This was interpreted as a dramatic change in shape using calculations based upon the Strutinsky shell-correction method~\cite{Frauendorf1975}.
The ground states of the heavier isotopes were calculated to be weakly deformed and oblate in nature, but when approaching the neutron mid-shell at $N=104$, this picture changed to a more-strongly deformed prolate shape.
These shapes are associated with structures based upon two different proton-hole excitations across the $Z=82$ shell closure, namely \mbox{$\pi$($0p$-$2h$)} and \mbox{$\pi$($2p$-$4h$)}.
Recently, the first direct evidence of shape coexistence in the even-mass Hg isotopes came from Coulomb-excitation experiments~\cite{Bree2014,Wrzosek-Lipska2015}, which quantified the deformation of ground and excited $0^{+}$ states for the first time in this region.

At $Z=82$, the lead isotopes remain spherical in their ground state all the way to mid-shell, as indicated by isotope-shift measurements employing laser spectroscopy~\cite{DeWitte2007,Seliverstov2009}.
In the case of $^{186}$Pb$_{104}$, competition between three shape minima is observed - oblate, prolate and the spherical ground state. This was inferred from $\alpha$-decay measurements of $^{190}$Po~\cite{Andreyev2000}, where the three states lying lowest in energy were observed to be $0^{+}$ states. 
This triple shape coexistence is apparent all around the mid-shell in the parabolic behaviour of the intruder energy levels as a function of mass number~\cite[Figure 3 of Ref.][]{Rahkila2010}, recently investigated down to $^{180}$Pb$_{98}$~\cite{Rahkila2010}.

The phenomenon persists in nuclei above $Z=82$, where the polonium isotopes were recently observed to have a much earlier and more gradual onset of deformation than observed in mercury~\cite{Cocolios2011}, without the unusual odd-even staggering~\cite{Seliverstov2013}.
One might consider that mercury ($Z=80$) and polonium ($Z=84$) are analogues with respect to their nucleon configuration; the oblate structure in the mercury isotopes, driven by $\pi$($0p$-$2h$) configurations, should manifest itself in polonium in $\pi$($2p$-$0h$) configurations and similarly for the prolate structure.
Indeed, the same parabolic behaviour of intruder states was observed when approaching mid-shell~\cite{Julin2001} and was interpreted to be of $\pi$($4p$-$2h$) configuration~\cite{Oros1999}.
Coulomb-excitation measurements recently determined multiple low-lying matrix elements for nuclei in the transitional region where the onset of deformation is observed~\cite{Kesteloot2015}.

\begin{figure}[!t]
\centering
\includegraphics[width=\columnwidth]{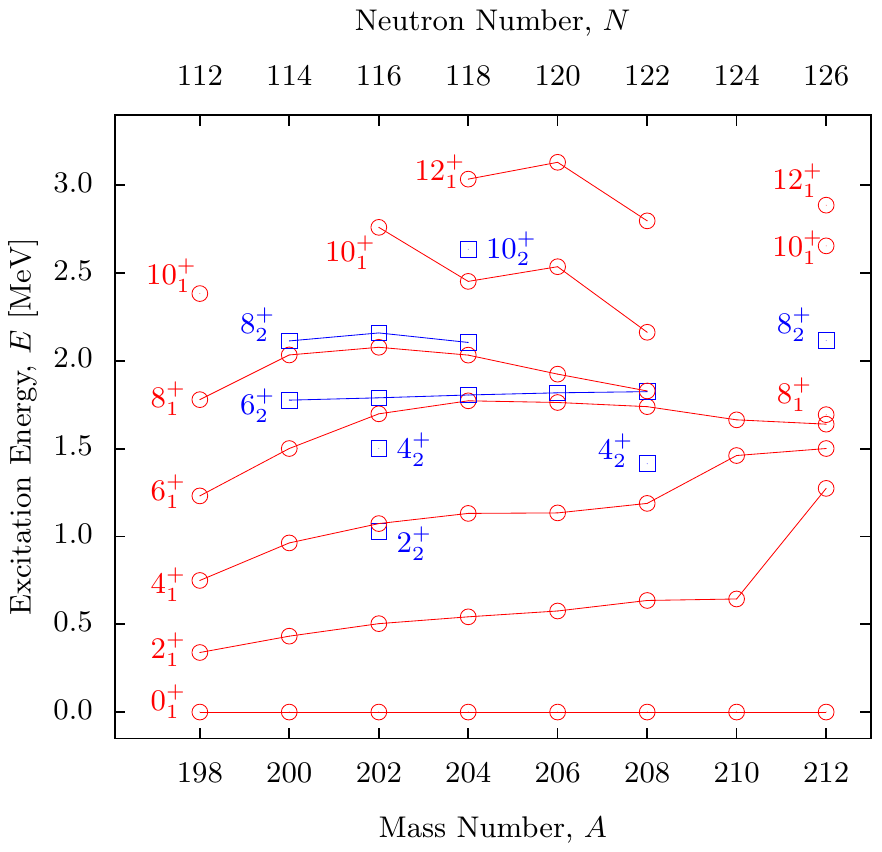}
\caption{(Color online) Systematics of excited states in even-even radon nuclei. The lowest-known even-spin positive-parity states are shown, with the first- (red) and second-excited (blue) states each connected by a solid line to guide the eye).}
\label{fig:system}  
\end{figure}

The radon isotopes ($Z=86$) can be expected to have similar proton-hole analogues to the platinum isotopes, where spectroscopic information on deformed intruder states exists beyond the neutron mid-shell~\cite{Dracoulis1994}. The energy level systematics of the even-spin positive-parity states in the light even-mass radon isotopes are shown in Figure~\ref{fig:system}, where one can observe decreasing excitation energy of the $2^{+}$ state towards $^{198}$Rn~\cite{Taylor1996,Taylor1999}.
A corresponding deviation from sphericity at $N=116$ is observed in the mean-square-charge radii~\cite{Borchers1987,Georg1995,*Georg1997}, earlier still than in the Po isotopes~\cite{Cocolios2011}. 
This may indicate that there is indeed a region of deformation towards the neutron mid-shell that is unreachable within the current experimental limitations.
A more detailed understanding, with complementary experimental probes, of the isotopes around this transition region, $^{198-204}$Rn($N=112-118$), would help to determine if this behaviour is in fact due to the presence of shape-coexisting intruder states.

\begin{figure}[!tb]
\centering
\includegraphics[width=0.98\columnwidth]{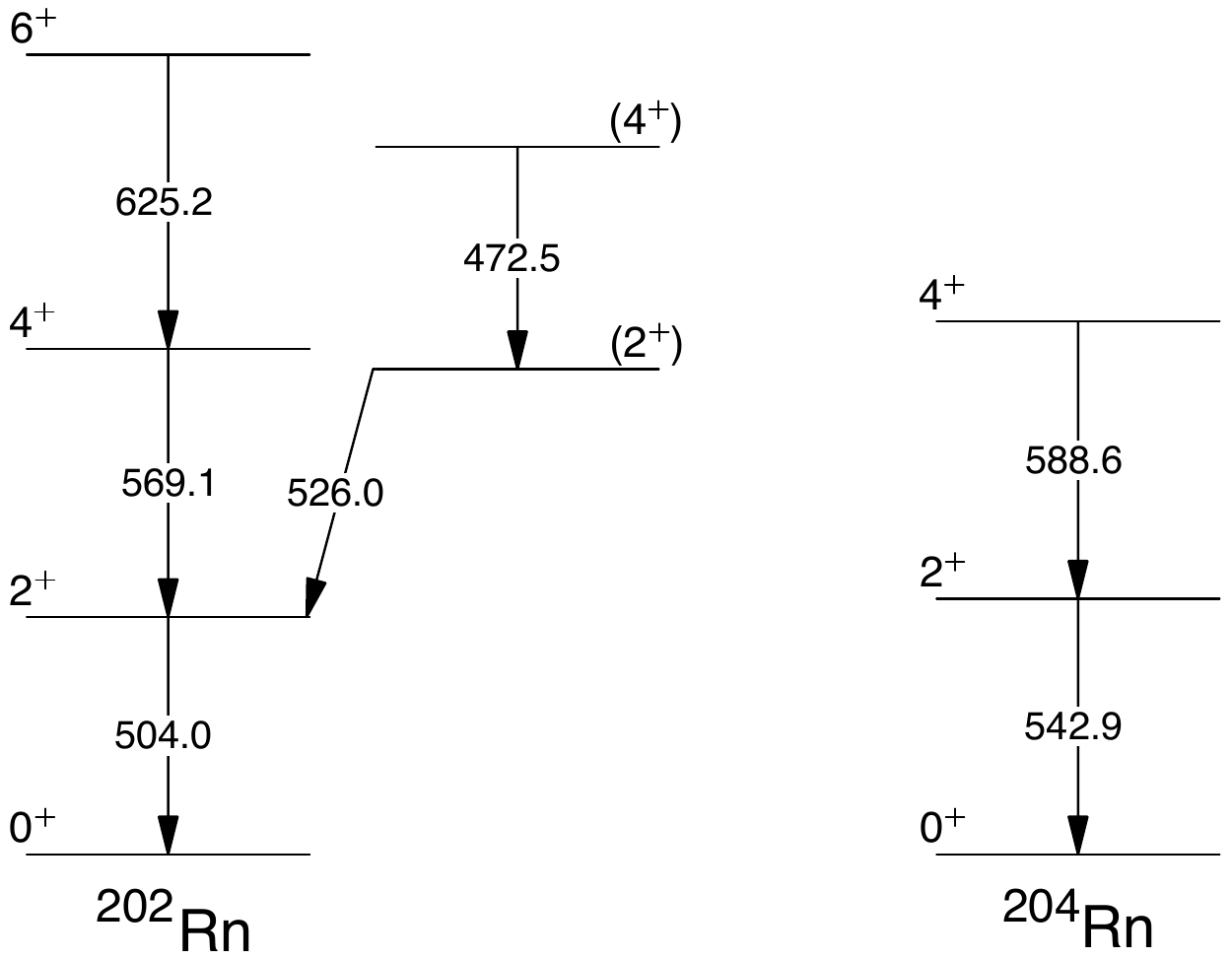}
\caption{Level schemes for $^{202}$Rn and $^{204}$Rn showing low-energy states included in the GOSIA analysis.}
\label{fig:levels}  
\end{figure}

Low-lying excited states in the isotopes around the $N=126$ shell closure are generally considered to be associated with a seniority scheme~\cite{Ressler2004,Grahn2013}, while lower masses are proposed to be candidates for vibrational nuclei.
The observed equal level spacing i.e. a ratio of the $4^{+}$ to $2^{+}$ excitation energy (R$_{42}$) close to 2, indicates a possible vibrational nature and the existence of a second $2^{+}$ state at a similar energy to the $4^{+}_{1}$ in $^{202}$Rn adds further weight to this argument.
The harmonic quadrupole vibrator should lead to a very definite and simple pattern of states with a single-phonon state with $I^{\pi}=2^{+}$, a triplet of two-phonon states with $J^{\pi} = 0^{+}$, $2^{+}$ and $4^{+}$, and so on.
As far as $^{202}$Rn and $^{204}$Rn~\cite{Dobson2002} are concerned, several of the expected members of vibrational multiplets are missing, although it is not presently clear if this is due to an experimental limitation. Their low-lying level schemes of interest to this study are shown in Figure~\ref{fig:levels}.
In particular, no observations of a excited $0^{+}$ states have been made in any of these nuclei.
Although its existence is expected in both a vibrational and intruder picture, the energy and $B(E2)$ values connecting $2^{+}$ states would definitively determine the structure.
Detailed investigations of excited $0^{+}$ states into the cadmium isotopes have proceeded in a similar vein~\cite{Garrett2008}, where the vibrational picture was found not to be adequate~\cite{Wood2012}.
Additionally, the presence of two near-parallel $6^{+}$ states in both nuclei is hard to accommodate in a simple vibrational picture.  

A detailed understanding of shape coexistence, or vibrational nuclei, will never be achievable from a single class of measurement.
A comprehensive picture of the underlying physics can only come from extraction of electromagnetic matrix elements involving a complementary set of experimental probes.
Transition matrix elements may be derived from lifetime measurements, in combination with precision branching and mixing ratios, from in-beam or decay spectroscopy.
Coulomb excitation allows not only the extraction of transition matrix elements but also of diagonal matrix elements, including their sign.
These can be used to further conclude on the sign of the spectroscopic quadrupole moment for excited states and hence, the type of nuclear deformation.
Multi-step Coulomb excitation needed to investigate low-lying non-yrast states in these nuclei, requires the availability of intense accelerated radioactive ISOL beams, which have only recently become available at facilities such as SPIRAL and REX-ISOLDE.
A pioneering example of this technique was in $^{74,76}$Kr~\cite{Clement2007} at SPIRAL. 
An intense program of Coulomb-excitation experiments at REX-ISOLDE has been underway to study the $Z=82$ region.
This facility is chosen as it is uniquely capable of producing beams of heavy proton-rich nuclei from spallation reactions.
Furthermore, key techniques such as laser ionisation have been developed to produce isobarically-pure secondary beams.
Experiments involving very heavy ($A>200$), post-accelerated beams have proven successful at REX-ISOLDE in recent years, including those employing radon~\cite{Gaffney2013}.
Studies such as these, performed at ISOL facilities around the world, are currently pushing the boundaries of nuclear spectroscopy on the precision frontier in exotic nuclei~\cite{Jenkins2014}.

In addition to the possibility of measuring electromagnetic matrix elements, Coulomb excitation is a well adapted technique for locating missing states, especially low-lying, non-yrast states that may not otherwise be populated in-decay or fusion evaporation experiments. Since low-lying $0^{+}$ states are key to the understanding of these nuclei, exploring the possibility of populating a $0^{+}_{2}$ state via a two-step Coulomb-excitation process is desirable.

\section{Experiment and Data Analysis}

Radioactive beams of $^{202}$Rn and $^{204}$Rn were produced at the ISOLDE facility in CERN via bombardment of a uranium-carbide primary target with 1.4-GeV protons from the PS Booster.
The target-ion-source coupling in this experiment was key to reduce isobaric impurities expected when working with a noble-gas beam.
A plasma ion source~\cite{Penescu2010} was utilised and an extraction voltage of 30~kV was applied along the transfer line and continuously cooled by a water flow in order to suppress the transport of less volatile elements.
At the beginning of the running period, the yield of the two radioactive species were measured using the dedicated ISOLDE tape station and found to be $9\times10^{5}$~ions/$\mu$C ($^{202}$Rn) and $2\times10^{7}$~ions/$\mu$C for $^{204}$Rn.
The singly-charged ions were accumulated and cooled in an ion trap, REX-TRAP~\cite{Wolf2003,Wenander2010}.
At intervals of 58~ms, the potential barrier was lowered allowing bunches of cooled ions to escape into an electron-beam ion source, REX-EBIS~\cite{Wolf2003,Wenander2010}, where the charge state of the ions was increased by charge breeding up to $47^{+}$.
The $^{202}$Rn and $^{204}$Rn beams were then accelerated to 2.9 and 2.845~MeV/u, in the 2008 and 2010 campaigns, respectively, by the REX linear accelerator~\cite{Voulot2008}.
A failure of the 9-gap resonator, the final element of the REX-LINAC, in the original 2008 campaign restricted the running period. This meant that a significant amount data, for both $^{204}$Rn and $^{202}$Rn, was taken at the lower beam energy of 2.28~MeV/$u$.

The secondary radioactive beams were incident on thin metallic foil targets positioned at the centre of the Miniball germanium detector array~\cite{Warr2013}.
The delivered beam currents at the target position were estimated to be around $3\times10^{4}$~ions/s for $^{202}$Rn and $2\times10^{5}$~ions/s for $^{204}$Rn.
The isobaric purity of the beam was monitored through inspection of the $\gamma$-ray spectrum obtained with a germanium detector positioned at the beam dump, approximately 3~m downstream of the target chamber.
Aside from transitions due to normal room background, the beam-dump spectrum only contained $\gamma$-ray transitions following the $\beta$- and $\alpha$-decay of the isotopes of interest.
However, during the second campaign in 2010, the cooling of the transfer line of the ion source failed.
This higher temperature allowed volatile elements to pass, in particular, a significant amount of the stable $^{202}$Hg, which caused contamination of the beam.
From the observation in the Coulomb-excitation spectrum of the $^{202}$Hg($2^{+}_{1} \rightarrow 0^{+}_{1}$) transition at 439.5~keV, in combination with the previously measured ${B(E2;2^{+}_{1} \rightarrow 0^{+}_{1})}$ value~\cite{Zhu2008}, the integrated beam current associated with $^{202}$Hg was deduced and represented 10\% of that associated with $^{202}$Rn.

\begin{figure}[!tb]
\centering
\def\svgwidth{1.0\columnwidth}
\includegraphics[width=\columnwidth]{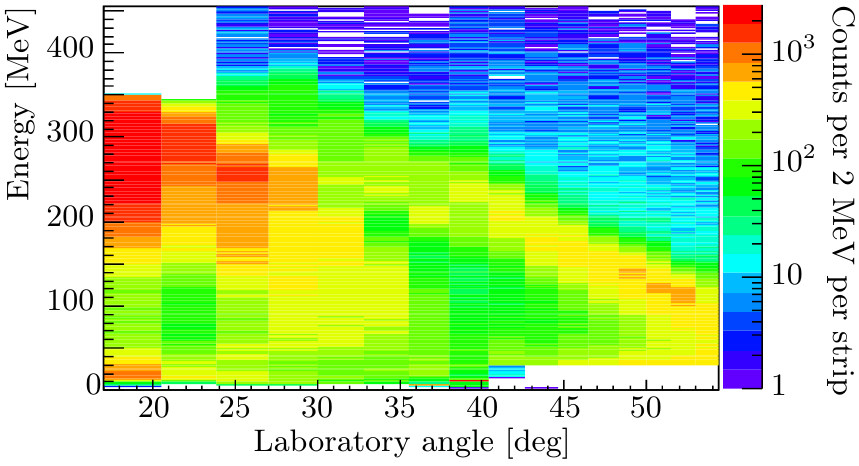}
\caption{(Color online) A two-dimensional spectrum, as a function of energy and laboratory angle, of scattered projectiles and recoils in the CD detector in the  $^{109}$Ag($^{202}$Rn) reaction, at a beam energy of 2.85~MeV/$u$ and target thickness of 1.9~mg/cm$^{2}$. There is no condition on the detection of a $\gamma$ ray.}
\label{fig:kinematics}  
\end{figure}

The Miniball array~\cite{Warr2013} comprises eight triple-cluster germanium detectors; each crystal is six-fold segmented, leading to a total of 144 discrete detector elements. The total efficiency of the array is $\approx7\%$ for 1.33~MeV $\gamma$ rays.
Scattered heavy ions were detected in an 500-$\mu$m-thick annular silicon double-sided silicon strip detector (DSSSD) segmented into four quadrants.
This CD detector, so-called due to its resemblance to a Compact-Disc, has 16 annular strips on the front face and 24 sectors on the back, and covered the range of laboratory angles from $\theta=16.2^{\circ}$ to $53.3^{\circ}$. 
Figure~\ref{fig:kinematics} illustrates the kinematics for scattering of $^{204}$Rn on $^{109}$Ag at a centre-of-target energy of 535~MeV.
The reactions are performed in inverse kinematics so there are two solutions for the projectile case. In addition, for the lowest laboratory angles, there is an ambiguity between the scattered projectile and target ions, such that the first two strips of the CD detector cannot be utilised in the analysis.
In order to resolve the issue of having an ambiguous conversion from laboratory angle to centre-of-mass (CoM) scattering angle, crucial for the calculation of the Coulomb-excitation cross sections, a coincidence gate on the recoiling target nuclei is applied. Here, the events corresponding to the second solution, at the very lowest CoM scattering angles, are not detected since the recoils do not have enough energy to exit the finite width of the target from the point of reaction.
Any of those that do (e.g. when the reaction occurs at the back of the target) are below the energy threshold of the CD detector.
Therefore, one can confidently assume that all recoil events are from the higher CoM-scattering-angle solution.

\begin{figure}[!tb]	
\centering
\includegraphics[width=\columnwidth]{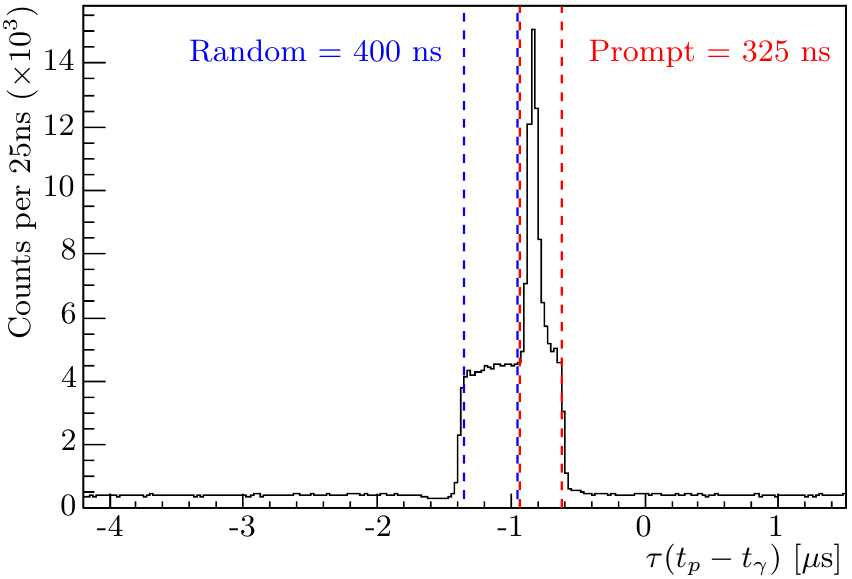}
\caption{Time difference between a $\gamma$ ray, which is acting as a trigger, and all correlated particles. The effects of the 800-ns downscaling window is clearly visible. Two regions are indicated in order to define a particle as in ``prompt'' coincidence, or in ``random'' coincidence.}
\label{fig:tdiff}  
\end{figure}

\begin{figure}[!tb]	
\centering
\includegraphics[width=\columnwidth]{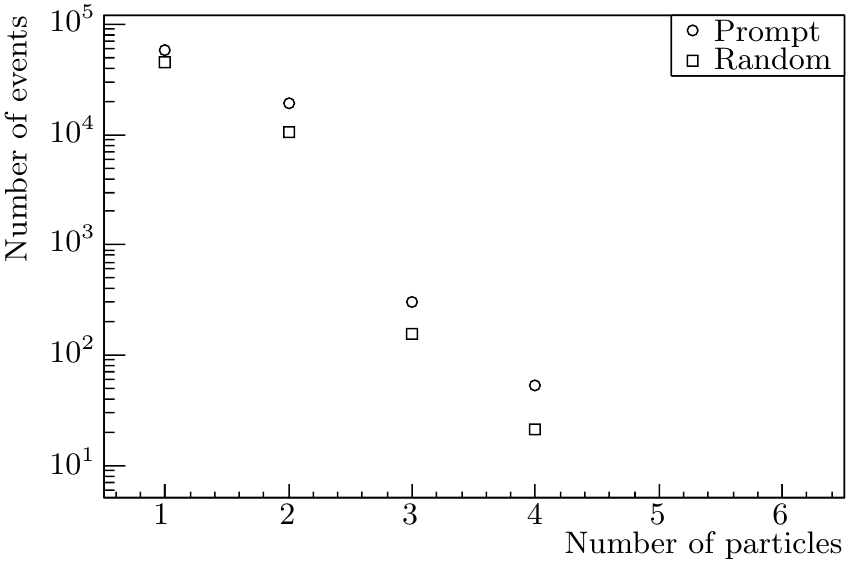}
\caption{Particle multiplicity for each $\gamma$-ray: the number of particles that are defined as prompt or random according to Figure~\ref{fig:tdiff}.}
\label{fig:multp}  
\end{figure}

\begin{figure}[!tb]
\centering
\includegraphics[width=\columnwidth]{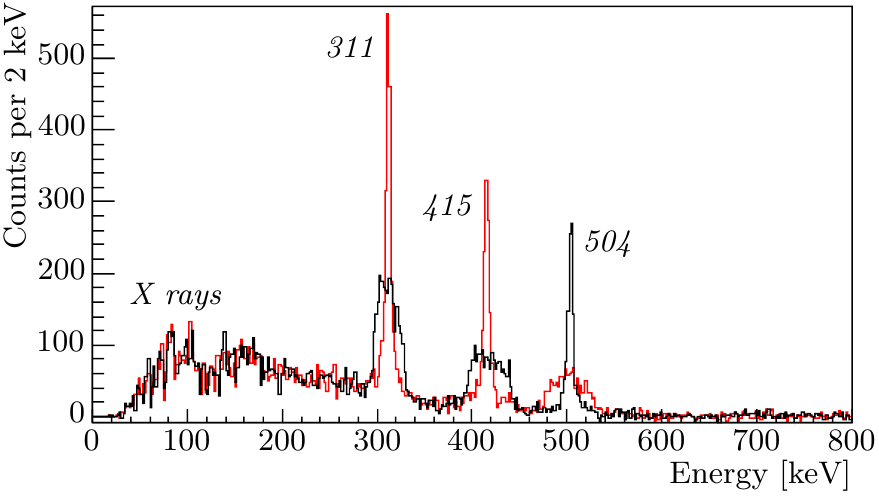}
\caption{(Color online) $\gamma$-ray de-excitation spectra associated with the Coulomb excitation of $^{202}$Rn on $^{109}$Ag at 2.90~MeV/u, Doppler-corrected for projectiles (black) and target recoils ({\color{red}red}). Only events identified in prompt coincidence with a recoiling target nucleus are shown; random events, with respect to the particle-$\gamma$ coincidence time, have been subtracted. Peaks are marked with their energy in keV.}
\label{fig:202onag}       
\end{figure}

\begin{figure}[!tb]
\centering
\includegraphics[width=\columnwidth]{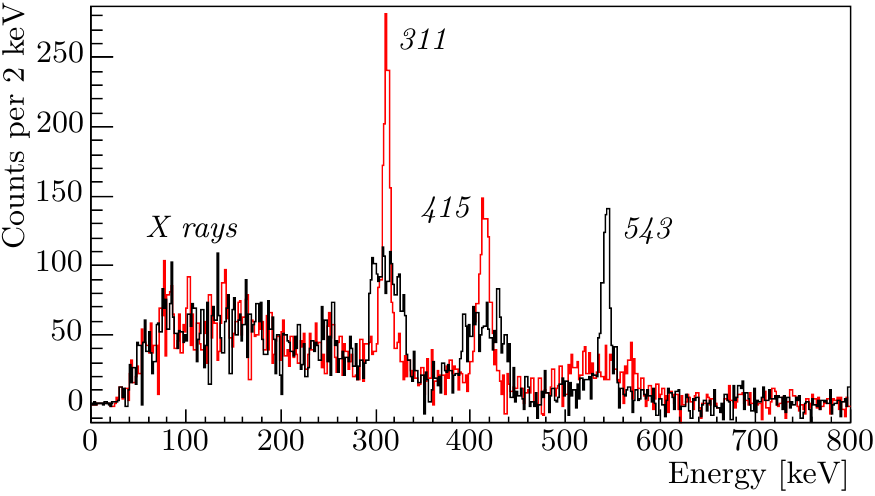}
\caption{(Color online) $\gamma$-ray de-excitation spectra associated with the Coulomb excitation of $^{204}$Rn on $^{109}$Ag at 2.90~MeV/u, Doppler-corrected for projectiles (black) and target recoils ({\color{red}red}). Only events identified in prompt coincidence with a recoiling target nucleus are shown.; random events, with respect to the particle-$\gamma$ coincidence time, have been subtracted. Peaks are marked with their energy in keV.}
\label{fig:204onag}  
\end{figure}

\begin{figure}[!tb]
\centering
\includegraphics[width=\columnwidth]{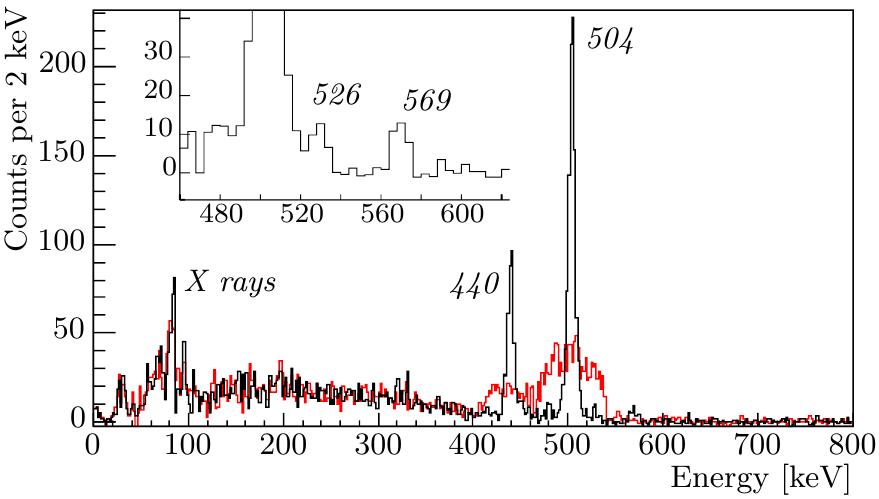}
\caption{(Color online) As in Figure~\ref{fig:202onag} but for the $^{120}$Sn target. The inset shows an expanded portion of the spectrum, with a bin width of 4~keV. Peaks are marked with their energy in keV.}
\label{fig:202onsn}       
\end{figure}

\begin{figure}[!tb]
\centering
\includegraphics[width=\columnwidth]{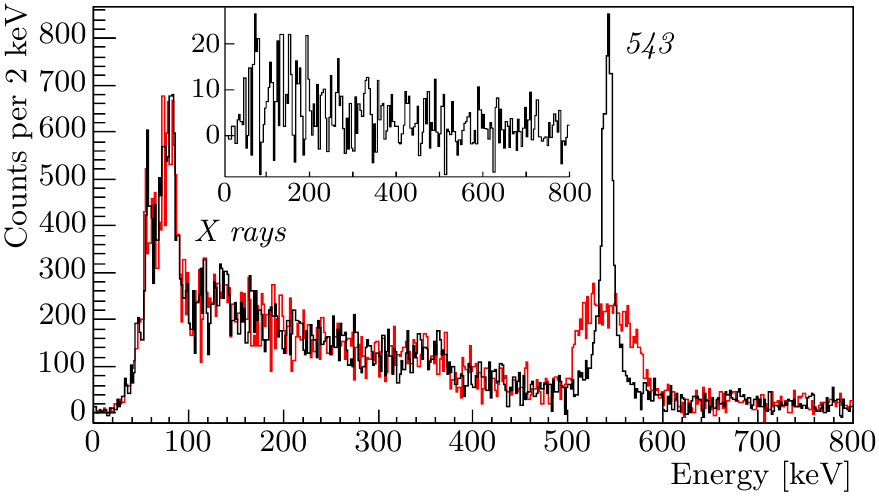}
\caption{(Color online) As in Figure~\ref{fig:204onag} but for the $^{120}$Sn target. Peaks are marked with their energy in keV. The inset shows the background-subtracted $\gamma$-$\gamma$ matrix, gated on the 543-keV $2^{+}_{1} \to 0^{+}_{1}$ transition, with a bin width of 4~keV.}
\label{fig:204onsn}       
\end{figure}

Triggered by the release of EBIS, data is collected from all detectors during 800-$\mu$s wide ``beam-on'' window following by an equally-wide ``beam-off'' window 4-10~ms later. In software, a correlation window of 6~$\mu$s is defined around the each $\gamma$-ray event of the ``beam-on'' window, and all particles that fall within this window are associated with that $\gamma$-ray.
In this way, it is possible for a single particle to be correlated to multiple $\gamma$ rays, but not vice versa.
Prompt and random windows are defined by taking the time difference between the particle and $\gamma$-ray triggers, as shown in Figure~\ref{fig:tdiff}.
The particle multiplicity, shown in Figure~\ref{fig:multp}, of each event can now be defined as $m$p-$n$r, where $m$($n$) is the number of prompt(random) particles.
In order to subtract the randomly coincident background from the spectra, 0p-2r and 0p-1r events are treated in exactly the same way as their prompt counterparts, but given a weight of $-T_{p}/T_{r}$, where $T_{p,r}$ is the width of the prompt and random time windows, respectively. 
The $\gamma$-ray spectra of Figures~\ref{fig:202onag}, \ref{fig:204onag}, \ref{fig:202onsn}, and~\ref{fig:204onsn} show background-subtracted 2p-0r events, where each particle can be identified as a recoil and projectile coincident within a particle-particle time window of 150~ns. 
Additionally, 1p-0r events, where only the recoil is identified, are included. In these cases, the projectile kinematics, i.e. laboratory angle and exit energy, must be reconstructed for the purposes of performing an optimal Doppler correction.
Here, two-body elastic scattering is assumed and the energy loss is calculated by integrating phenomenological stopping power curves fitted to data from SRIM~\cite{srim2010}.
By definition, the $\gamma$-ray multiplicity is one, but $\gamma$-$\gamma$ events can still be built by looking for events that have the same prompt particle correlations.
In these experiments, such $\gamma$-$\gamma$ coincidences didn't provide any additional information (see inset of Figure~\ref{fig:204onsn}).
As shown in Figure~\ref{fig:multp}, higher-order particle multiplicities do not account for a significant fraction of the data and are not taken into account in this analysis. Events where both $m$ and $n$ are greater than zero, i.e., at least one prompt particle and at least one random particle, are also not considered due to ambiguity in assigning prompt or random status.
In the case that this represents a significant amount of data, it is possible to assume a prompt nature for such events, but the weighting of random events must be re-considered to account for this.
Usually, the ratio of intensities of transitions associated with $\beta$-decaying daughter products of the beam, assumed to be purely random in time, is then used.

Due to the inherent difficulties in performing an absolute normalisation to elastically-scattered particles with Miniball, caused by an imprecise knowledge of the dead-time with different coincidence conditions, normalisation to the excitation of the target is preferred~\cite{Zielinska2015}.
In this case, the $^{202}$Rn($^{204}$Rn) beam was incident on a 4.0(1.9)~mg/cm$^{2}$ target of $^{109}$Ag, for which the relevant matrix elements are sufficiently well-established experimentally.
The resulting de-excitation $\gamma$-ray spectra are shown in Figures~\ref{fig:202onag}(\ref{fig:204onag}).
In the excitation process on the $^{109}$Ag target, only the $2^{+}_{1}$ states are populated in the $^{202,204}$Rn projectiles.
This means that the $B(E2;2^{+}_{1}\rightarrow0^{+}_{1})$ and $Q_{s}(2^{+}_{1})$ can be determined by utilising the first-order assumption that matrix elements connecting higher-lying states, of which we have no direct experimental information, do not contribute.

Due to the presence of de-excitation $\gamma$-rays from the target that are Doppler shifted differently to those from the projectile, it is not easy to locate weak $\gamma$-ray transitions in the projectile.
Accordingly, data was also taken on a 2.0~mg/cm$^{2}$ target of $^{120}$Sn, chosen to reduce the number and intensity of $\gamma$-ray transitions resulting from target excitation.
A high-lying first-excited $2^{+}$ state at 1171~keV, with a relatively small $B(E2)$ of 11.4~W.u., means that this state is not strongly populated.
This reduces the complexity of the spectra as well as the background from Compton-scattered, escaped events, as can be seen in Figures~\ref{fig:202onsn} and~\ref{fig:204onsn}.
In all of the $\gamma$-ray spectra, the intensity of radon K X-rays is markedly high, inconsistent with the expectations of internal conversion of $E2$ transitions. The residual fraction of these X rays is associated to K-vacancy creation in atomic collisions between the high-Z beam and target~\cite{Bree2015}.
Population of higher-lying states in $^{204}$Rn was inconclusive based on the $^{120}$Sn-target data (see Figure~\ref{fig:204onsn}), possibly due in part to a poor $\gamma$-ray resolution caused by noise on the CD detector in this part of the experiment, which affected the Doppler correction.
It may also be that the population of the states was simply below the detection limit of the experiment. An upper limit for the observation of the $4^{+}_{1} \rightarrow 2^{+}_{1}$ transition was determined.
In $^{202}$Rn, the $4^{+}_{1}$ state is a little lower in energy and there exists a previously-observed $2^{+}_{2}$ state at 1029~keV. Both of these states are clearly populated in the $^{120}$Sn-target data (see Figure~\ref{fig:202onsn}), albeit with low intensity.
It helps that both transitions sit at higher energy in the spectra than the dominant $2^{+}_{1} \rightarrow 0^{+}_{1}$ transitions, since they are clear of the Compton background and can be fitted with a smaller uncertainty.
The extracted intensities are presented in Tables~\ref{tab:yields_202} and~\ref{tab:yields_204}. No additional data is obtained from the lower-beam-energy runs and it is not considered in the cross-section analysis due to large uncertainties on $\gamma$-ray intensities.

\begin{table}[tb]
\begin{center}
\caption{Intensities of $\gamma$-ray transitions observed in the current Coulomb-excitation experiments of $^{202}$Rn. Efficiency correction has been performed, relative to the $2^{+}_{1} \rightarrow 0^{+}_{1}$ transition of the projectile in each experiment. Transition intensities in the $^{109}$Ag target are also included and can be identified by the odd-spin transitions.}
\label{tab:yields_202}
\begin{ruledtabular}
\begin{tabular}{cccc}
Beam Energy & Target & Transition  & $I_{\gamma}$ \\[2pt]
\hline
2.845~MeV/$u$ &	$^{120}$Sn			& $2^{+}_{1} \rightarrow 0^{+}_{1}$ & 990(37) \\
 &			(2.0~mg/cm$^{2}$)	& $2^{+}_{2} \rightarrow 2^{+}_{1}$ & 27(7) \\
 &		  					& $4^{+}_{1} \rightarrow 2^{+}_{1}$ & 29(6) \\
 &							& $2^{+}_{2} \rightarrow 0^{+}_{1}$ & $<19(9)$ \\[2pt]
2.90~MeV/$u$ &	$^{109}$Ag			& $2^{+}_{1} \rightarrow 0^{+}_{1}$ & 923(40) \\
 & 			(4.0~mg/cm$^{2}$)	& $3/2^{-}_{1} \rightarrow 1/2^{-}_{1}$ & 1260(60) \\
 & 							& $5/2^{-}_{1} \rightarrow 1/2^{-}_{1}$ & 1000(50) \\
\end{tabular}
\end{ruledtabular}
\end{center}
\end{table}

\begin{table}[tb]
\begin{center}
\caption{Intensities of $\gamma$-ray transitions observed in the current Coulomb-excitation experiments of $^{204}$Rn. Efficiency correction has been performed, relative to the $2^{+}_{1} \rightarrow 0^{+}_{1}$ transition of the projectile in each experiment. Transition intensities in the $^{109}$Ag target are also included and can be identified by the odd-spin transitions}
\label{tab:yields_204}
\begin{ruledtabular}
\begin{tabular}{cccc}
Beam Energy & Target & Transition  & $I_{\gamma}$ \\[2pt]
\hline
2.845~MeV/$u$ & 	$^{120}$Sn			& $2^{+}_{1} \rightarrow 0^{+}_{1}$ & $6130(200)$ \\
 &			(2.0~mg/cm$^{2}$)	& $2^{+}_{2} \rightarrow 2^{+}_{1}$ & $<190(160)$ \\
 &		  					& $4^{+}_{1} \rightarrow 2^{+}_{1}$ & $<240(90)$ \\[2pt]
2.90~MeV/$u$ & 	$^{109}$Ag			& $2^{+}_{1} \rightarrow 0^{+}_{1}$ & 660(40) \\
 & 			(1.9~mg/cm$^{2}$)	& $3/2^{-}_{1} \rightarrow 1/2^{-}_{1}$ & 720(50) \\
 & 							& $5/2^{-}_{1} \rightarrow 1/2^{-}_{1}$ & 700(50) \\
\end{tabular}
\end{ruledtabular}
\end{center}
\end{table}

Aside from the known $4^{+}_{1}$ state in $^{204}$Rn, there is the potential for the population of an unobserved $2^{+}_{2}$ state. Assuming it decays predominantly to the $2^{+}_{1}$ state as in $^{202}$Rn, the $\gamma$-ray de-excitation could form a doublet with the ${2^{+}_{1} \to 0^{+}_{1}}$ transition at 543~keV.
This would place the state around twice the energy of the $2^{+}_{1}$ state, something that is expected with a vibrational-like structure.
To investigate this possibility, the background-subtracted $\gamma$-$\gamma$ matrix for the $^{120}$Sn-target data was projected with a gate between 520 and 570~keV, as shown in the inset of Figure~\ref{fig:204onsn}.
A 1$\sigma$ upper limit of a peak~\cite{Helene1983} was determined for the region between 520 and 570~keV of 21(19) counts.
The $\gamma$-$\gamma$ efficiency, $\epsilon_{\gamma\gamma}(E_{\gamma})$, was determined at 311~keV through the $5/2^{-}_{1} \to 3/2^{-}_{1} \to 1/2^{-}_{1}$ cascade in $^{109}$Ag and extrapolated using the singles efficiency determined for a $^{152}$Eu/$^{133}$Ba source combination to give $\epsilon_{\gamma\gamma}(543~\mathrm{keV})=11(3)\%$.
Consequently, the 1$\sigma$ upper limit of the number of counts in the singles spectrum is 190(160) counts, which corresponds to less than 3\% of $I_{\gamma}(2^{+}_{1} \to 0^{+}_{1})$ transition.
Assuming a similar excitation probability for the Ag ($Z=47$) target as the Sn ($Z=50$) target, one can assume that this would not significantly affect the determination of the $B(E2; 2^{+}_{1} \to 0^{+}_{1})$ value, since it is less than the statistical precision of the transition intensity.

\section{Results}

For the Coulomb-excitation analysis, the \gosia{}~\cite{Czosnyka1983,GosiaManual} code was utilised in order to calculate excitation probabilities, and consequently de-excitation $\gamma$-ray intensities, for a given set of electromagnetic matrix elements.
The calculated intensities are then compared to experimental data, along with additional spectroscopic information, such as excited-state lifetimes, $E2/M1$ mixing ratios and $\gamma$-ray branching ratios.
Conversion coefficients used in \gosia{} were calculated using the BrIcc data tables~\cite{Kibedi2008}.
A $\chi^{2}$-like, least-squares function is constructed and can be minimised with respect to the electromagnetic matrix elements as input parameters, along with a set of normalisation constants.
For the cases where normalization to the $^{109}$Ag target excitation was used, a special version of the code, \gosia{}2, is employed.
Here, the total $\chi^{2}$ is calculated for fixed values of the projectile matrix elements, ${\langle 0^{+}_{1} \| E2 \| 2^{+}_{1} \rangle}$ and ${\langle 2^{+}_{1} \| E2 \| 2^{+}_{1} \rangle}$, scanning a large-scale two-dimensional surface in order to search for the best solution at $\chi^{2}_{\mathrm{min}}$.
The associated $1\sigma$ uncertainties can then be extracted by cutting the surface at $\chi^{2}_{\mathrm{min}}+1$ and projecting the limits to the relevant axis.
These procedures are described in detail in Ref.~\cite{Zielinska2015}.

In the first step, the level schemes as shown in Figure~\ref{fig:levels} are defined in \gosia{}2, where the $6^{+}_{1}$ and $4^{+}_{2}$ are buffer states in $^{202}$Rn ($4^{+}_{1}$ in $^{204}$Rn) to prevent an artificial build-up of population in the highest-energy observed states.
Both the $^{202}$Rn and $^{204}$Rn data are segmented into five different angular ranges, utilising the segmentation of the CD detector, yielding five independent experiments.
This gives a total of five data points in the projectile system; the intensity ($I_{\gamma}$) of the $2^{+}_{1} \rightarrow 0^{+}_{1}$ transition in each experiment.
There are, however, seven parameters; the matrix elements ${\langle 0^{+}_{1} \| E2 \| 2^{+}_{1} \rangle}$ and ${\langle 2^{+}_{1} \| E2 \| 2^{+}_{1} \rangle}$, plus five normalisation constants, which can be considered as a product of the integrated beam current, live-time fraction and particle-$\gamma$ efficiencies at the given scattering angle.
The target system is over-determined with ten transition intensities, two in each of the five independent experiments (angular ranges), in addition to the nine additional spectroscopic data presented in Table~\ref{tab:specdata109ag}, fitted to a total of seven matrix elements and five normalisation constants.
These five normalisation constants are shared between the projectile and target systems and can be fitted simultaneously in both data sets.
This allows for an over-determination of the whole system, which can be reduced to a two-parameter system with five data points for the projectile. 

\begin{table}[tb]
\begin{center}
\caption{Spectroscopic data related to the low-lying level-scheme ($1/2^{-}_{1}$, $3/2^{-}_{1}$, $5/2^{-}_{1}$) of $^{109}$Ag included in the \gosia{}2 fit. An average was taken of the two possible solutions for ${\langle 5/2^{-}_{1} \| E2 \| 5/2^{-}_{1} \rangle}$. Matrix elements connecting the higher-lying states ($3/2^{-}_{2}$, $5/2^{-}_{2}$) were determined from previous Coulomb-excitation measurements~\cite{Robinson1970,Zielinska2009a} and fixed in the fit. The $9/2^{+}$ isomeric state was not included.}
\label{tab:specdata109ag}
\begin{ruledtabular}
\begin{tabular}{ccc}
\multicolumn{2}{c}{Spectroscopic data for $^{109}$Ag}	&	Reference \\[1pt]
\hline
$B(E2;1/2^{-}_{1}\rightarrow 3/2^{-}_{1})$	 &  $0.222(19)$~$e^{2}$b$^{2}$  &  \cite{Robinson1970} \\[2pt]
$B(E2;1/2^{-}_{1}\rightarrow 5/2^{-}_{1})$	 &  $0.320(26)$~$e^{2}$b$^{2}$  &  \cite{Robinson1970} \\[2pt]
${\langle 3/2^{-}_{1} \| E2 \| 3/2^{-}_{1} \rangle}$	 & $-1.3^{+0.3}_{-0.4}$~$e$b  & \cite{Zielinska2009a} \\[2pt]
${\langle 5/2^{-}_{1} \| E2 \| 5/2^{-}_{1} \rangle}$	 & $-0.21$ or $-0.56$~$e$b  &  \cite{Throop1972} \\[2pt]
$\frac{I_{\gamma}(5/2^{-}_{1} \rightarrow 3/2^{-}_{1})}{I_{\gamma}(5/2^{-}_{1} \rightarrow 1/2^{-}_{1})}$ & $0.069(16)$  &  \cite{Blachot2006} \\[2pt]
$\tau(3/2^{-}_{1})$  &  $8.5(10)$~ps  &  \cite{Miller1974} \\[2pt]
$\tau(5/2^{-}_{1})$  &  $47(2)$~ps  &  \cite{Loiselet1989} \\[2pt]
$\delta(3/2^{-}_{1} \rightarrow 1/2^{-}_{1})$ & $-0.196(27)$  &  \cite{Robinson1970} \\[2pt]
$\delta(5/2^{-}_{1} \rightarrow 3/2^{-}_{1})$ & $-0.039(17)$  &  \cite{Robinson1970} \\[2pt]
\end{tabular}
\end{ruledtabular}
\end{center}
\end{table}

\begin{figure}[!t]
\centering
\includegraphics[width=\columnwidth]{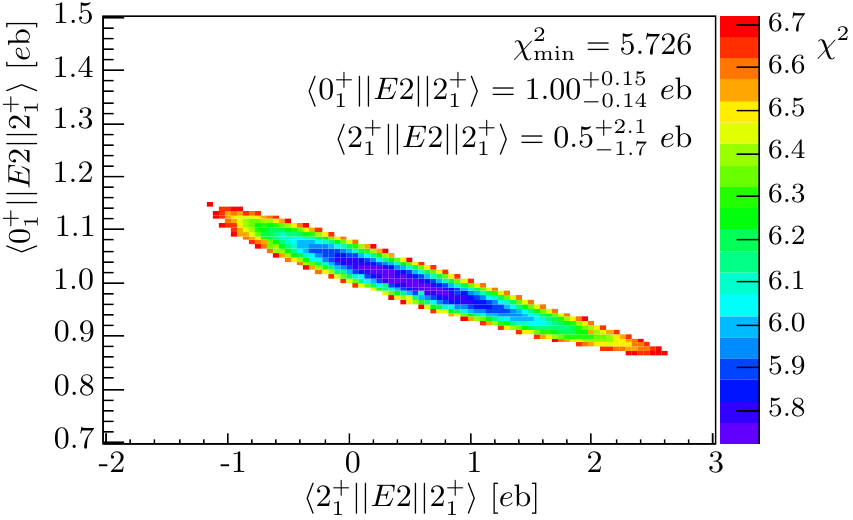}
\caption{(Color online) Two-dimensional total-$\chi^{2}$ surface for $^{202}$Rn on $^{109}$Ag at 2.9~MeV/u, extracted from \gosia{}. The data were segmented in to five angular ranges.}
\label{fig:chisqsurface202}  
\end{figure}

\begin{figure}[!t]
\centering
\includegraphics[width=\columnwidth]{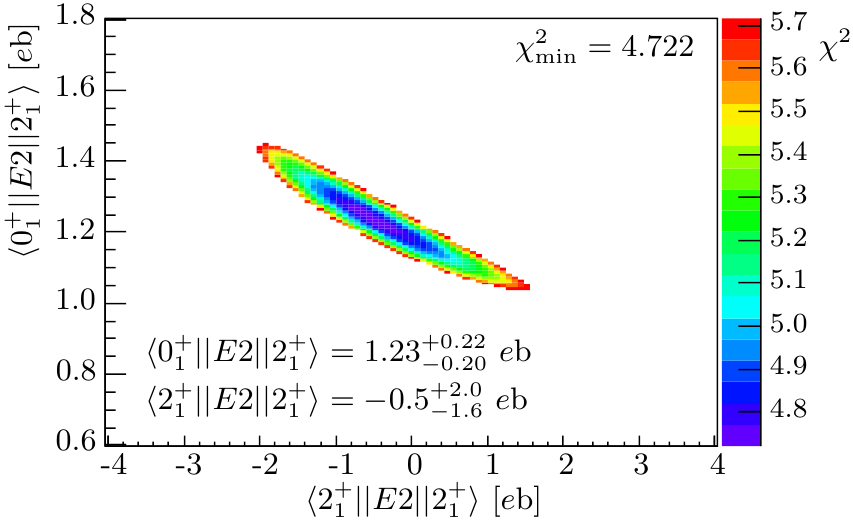}
\caption{(Color online) Two-dimensional total-$\chi^{2}$ surface for $^{204}$Rn on $^{109}$Ag at 2.9~MeV/u, extracted from \gosia{}. The data were segmented in to five angular ranges.}
\label{fig:chisqsurface204}  
\end{figure}

Two-dimensional $\chi^{2}$ surfaces are plotted in Figures~\ref{fig:chisqsurface202} and~\ref{fig:chisqsurface204} for $^{202}$Rn and $^{204}$Rn, respectively.
The kinematics of the experimental set-up limited the observed range of CoM scattering angles, which in turn limited the sensitivity to the diagonal matrix element, ${\langle 2^{+}_{1} \| E2 \| 2^{+}_{1} \rangle}$.
A good determination of the spectroscopic quadrupole moment, $Q_{s}(2^{+}_{1})$, which is proportional to ${\langle 2^{+}_{1} \| E2 \| 2^{+}_{1} \rangle}$, requires not only statistical precision but data at both high and low scattering angles.
This in turn achieves a variation in sensitivity to subtle higher-order effects.
For the current data set, a strong overlap in the $\chi^{2}$ functions of the different experiments leads to an elongation of the 1$\sigma$ confidence region in the ${\langle 2^{+}_{1} \| E2 \| 2^{+}_{1} \rangle}$ axis.
The strong correlation between the two parameters means that the determination of ${\langle 0^{+}_{1} \| E2 \| 2^{+}_{1} \rangle}$ or ${B(E2;2^{+}_{1} \rightarrow 0^{+}_{1})}$ is also adversely affected, increasing the projected uncertainty.
Under the assumption of no second-order effect for $Q_{s}$, the uncertainty is equivalent to that of the statistical uncertainty of the $\gamma$-ray intensity, but under-estimates the true uncertainty by a factor $\simeq3.5$ in both $^{202,204}$Rn.

Following the extraction of $B(E2;2^{+}_{1}\rightarrow0^{+}_{1})$, the higher-statistics data for Coulomb excitation on the $^{120}$Sn target is analysed in a second step.
The ratio of transition intensities ${I_{\gamma}(4^{+}_{1} \rightarrow 2^{+}_{1})/I_{\gamma}(2^{+}_{1} \rightarrow 0^{+}_{1})}$ can be considered to be almost directly proportional to the $B(E2;4^{+}_{1} \rightarrow 2^{+}_{1})$ value, with negligible influence from other matrix elements, including ${\langle 0^{+}_{1} \| E2 \| 2^{+}_{1} \rangle}$.
This is because the population of the $4^{+}_{1}$ state occurs almost exclusively in the two-step $E2$-excitation process involving the $2^{+}_{1}$ state.
The population of the $2^{+}_{1}$ is known very well from $I_{\gamma}(2^{+}_{1} \rightarrow 0^{+}_{1})$ since any significant feeding (i.e. from $4^{+}_{1}$ and $2^{+}_{2}$) can be accounted for. In $^{204}$Rn, the upper limit of $I_{\gamma}(4^{+}_{1} \rightarrow 2^{+}_{1})$ is used to calculate an upper limit for $B(E2;4^{+}_{1} \rightarrow 2^{+}_{1})$, shown in Table~\ref{tab:me}.

In the current experiment, the population of the $2^{+}_{2}$ state in $^{202}$Rn can be considered to occur exclusively via a two-step $E2$ excitation via the $2^{+}_{1}$ state.
The single-step process, directly from the ground state, can be assumed to be negligible due to the combination of the large energy difference and the small $B(E2; 2^{+}_{2} \rightarrow 0^{+}_{1})$ relative to the $B(E2; 2^{+}_{2} \rightarrow 2^{+}_{1})$ extracted from the upper limit of the branching ratio of 6.9\%~\cite{Gillespie_privcomms}.
Additionally, $M1$ excitation is calculated to be more than 100 times weaker than the corresponding $E2$ between the two $2^{+}$ states.
No complementary data (such as the lifetime, $\tau_{2^{+}_{2}}$, $E2/M1$ mixing ratio, $\delta(2^{+}_{2} \rightarrow 2^{+}_{1})$ or conversion coefficient, $\alpha(2^{+}_{2} \rightarrow 2^{+}_{1})$)  are available to constrain the ${\langle 2^{+}_{1} \| M1 \| 2^{+}_{2} \rangle}$ matrix element and consequently it is currently not possible to extract its value.
It can however be shown that the current data set is insensitive to the value of the $M1$ component, and the determination of ${\langle 2^{+}_{1} \| E2 \| 2^{+}_{2} \rangle}$ is unaffected.
The $M1$ matrix element was coupled to the $E2$ matrix element using $|\delta(2^{+}_{2} \rightarrow 2^{+}_{1})|=1.1$, by comparison to known values in the region.

All of the data for $^{202}$Rn, collected with both $^{109}$Ag and $^{120}$Sn targets, are fitted using the least-squares search code, \gosia{}~\cite{Czosnyka1983,GosiaManual}, in order to fully investigate all potential couplings to unknown matrix elements~\cite{Zielinska2015}. In the final fit, many matrix elements were coupled, or fixed to reasonable values, when the fit was found to be insensitive to their values. The diagonal $E2$ matrix elements of the $4^{+}_{1}$ and $2^{+}_{2}$ were coupled to their transitional counterparts, assuming a constant $Q_{0}$ and $K=0$ within the rigid rotor model.
A particular concern with regards to correlations is the ${\langle 4^{+}_{1} \| E2 \| 2^{+}_{2} \rangle}$ matrix element, which influences the populations of both the $4^{+}_{1}$ and $2^{+}_{2}$ states.
It was fixed to 0.005~$e$b in the final fit, although values up to 1.5~$e$b were tested and shown to influence the final result at the few percent level, much less than the statistical uncertainty.
For the correlated error calculation, it was allowed to vary with limits of $\pm1.5$~$e$b.

Once the $\chi^{2}$ minimum is found, the uncertainties are calculated by \gosia{} in a two-stage process.
At this point, all couplings and fixed matrix elements are freed in order to correctly include the influence of correlations to unknown matrix elements.
Firstly, the diagonal, or uncorrelated, uncertainties on each matrix element are computed by varying it about the minimum until an increase in $\chi^{2}$ is achieved, satisfying the $1\sigma$ condition~\cite{GosiaManual}.
At the same time, a multi-dimensional correlation matrix is built, which is then used in the second step in order to compute the fully correlated errors on each matrix element.
It was shown that the ${\langle 4^{+}_{1} \| E2 \| 2^{+}_{1} \rangle}$ matrix element is insensitive to changes in other transitional matrix elements and only very weakly ($<<1\sigma$) dependent on ${\langle 4^{+}_{1} \| E2 \| 4^{+}_{1} \rangle}$.
This leads to an uncertainty on $B(E2;4^{+}_{1} \rightarrow 2^{+}_{1})$ roughly equivalent to the statistical uncertainty of $I_{\gamma}(4^{+}_{1} \rightarrow 2^{+}_{1})$.
For ${\langle 2^{+}_{1} \| E2 \| 2^{+}_{2} \rangle}$, however, the correlations play a much stronger role and the uncertainty on $B(E2;2^{+}_{2} \rightarrow 2^{+}_{1})$ is relatively large (see Table~\ref{tab:me}).

\begin{table}[tb]
\begin{center}
\caption{Transition strengths, $B(E2)$, and spectroscopic quadrupole moments, $Q_{s}$, along with their uncertainties obtained from the two-dimensional $\chi^{2}$ analyses and \gosia{} minimisation. In the case of $^{202}$Rn, the final values are extracted from the full simultaneous analysis of data on both the $^{109}$Ag and $^{120}$Sn targets. The uncertainties include correlations to all seven matrix elements in the fit. The fit is shown to converge with the two-dimensional $\chi^{2}$ analysis and produces consistent uncertainties, proving that the correlations are small.}
\label{tab:me}
\begin{ruledtabular}
\begin{tabular}{ccc}
 & $^{202}$Rn  & $^{204}$Rn \\[2pt]
\hline
${B(E2;2^{+}_{1} \rightarrow 0^{+}_{1})}$ & $29^{+8}_{-8}$~W.u.  &  $43^{+17}_{-12}$~W.u. \\
${B(E2;2^{+}_{2} \rightarrow 2^{+}_{1})}$ & $160^{+90}_{-50}$~W.u.  & -- \\
${B(E2;2^{+}_{2} \rightarrow 0^{+}_{1})}$ & $<0.4(3)$~W.u.  & -- \\
${B(E2;4^{+}_{1} \rightarrow 2^{+}_{1})}$ & $63^{+18}_{-18}$~W.u.  & $<74(30)$ \\
$Q_{s}(2^{+}_{1})$  & $0.9^{+2.9}_{-1.8}$~$e$b  &  $-0.4^{+1.5}_{-1.2}$~$e$b \\
\end{tabular}
\end{ruledtabular}
\end{center}
\end{table}

\section{Discussion}

\begin{figure}[tb]
\centering
\includegraphics[width=\columnwidth]{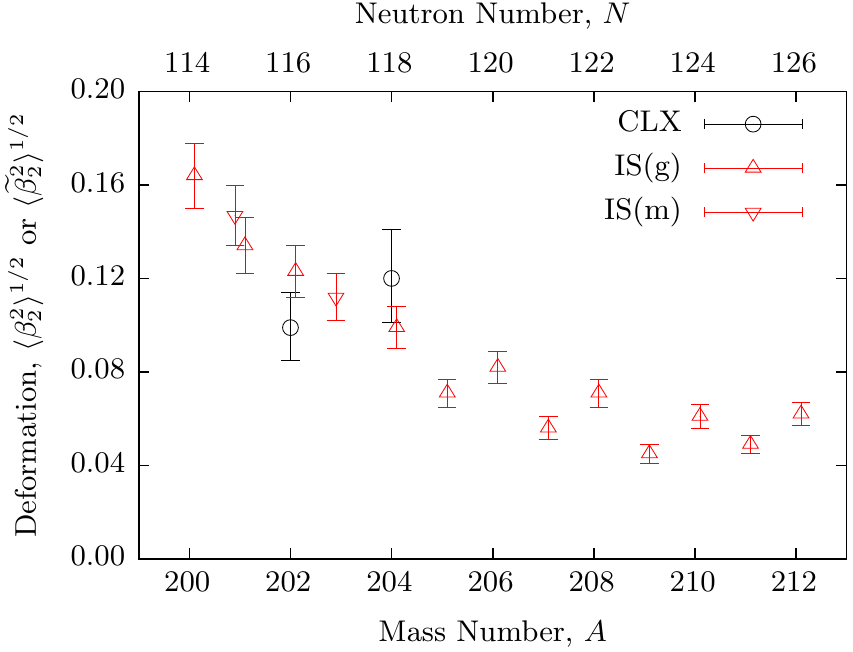}
\caption{(Color online) Experimental $\langle\beta_{2}^{2}\rangle^{1/2}$ values deduced from the $B(E2; 0^{+}_{1} \to 2^{+}_{1})$ values measured in this work (black circles, ``CLX'') and those from isotope-shift measurements and liquid-drop model for both the ground ({\color{red}red} down triangles, ``IS(g)'') and isomeric  ({\color{red}red} up triangles, ``IS(m)'') states. The uncertainties on the latter are dominated by the propagation of the uncertainty in the Grodzins-Raman rule~\cite{Raman2001}, which is a systematic contribution. The isotope shift values are slightly offset from integer $A$ values for clarity of presentation.}
\label{fig:deformation}  
\end{figure}

Under the assumption that the quadrupole charge distribution is uniform and can describe the nuclear shape, the deformation can be deduced from the following sum over $B(E2)$ values~\cite{Raman2001}:
\begin{equation} \label{eq:beta2_clx}
\sum_{i} B(E2; 0^{+}_{1} \to 2^{+}_{i}) = \left( \frac{3}{4\pi} Z e R_{0}^{2} \right)^{2} \langle\beta_{2}^{2}\rangle	~,
\end{equation}
where $Ze$ is the nuclear charge and $R_{0}=1.2A^{1/3}$~fm.
From the limit of the $2^{+}_{2}$ branching ratio in $^{202}$Rn~\cite{Gillespie_privcomms} (see Table~\ref{tab:me}) it is a reasonable assumption that the $E2$ transition strength from the ground state is dominated by the first-excited $2^{+}$ state in these nuclei.
Thus, one can limit the sum in Equation~\ref{eq:beta2_clx} to $i=1$.
The deduced deformations are then ${\langle\beta_{2}^{2}\rangle^{1/2}(^{202}\mathrm{Rn}) = 0.099^{+0.015}_{-0.014}}$ and ${\langle\beta_{2}^{2}\rangle^{1/2}(^{204}\mathrm{Rn}) = 0.120^{+0.021}_{-0.019}}$, indicating a weak deformation.
Another indication of the ground-state deformation can be deduced from isotope-shift measurements~\cite{Borchers1987,Georg1995,*Georg1997}, where mean-square charge radii, $\langle r^{2} \rangle_{A}$, is related to the deformation (to first order) in the following way:
\begin{equation} \label{eq:beta2_laser}
\langle r^{2} \rangle_{A} \approx \langle r^{2} \rangle_{A}^{\mathrm{sph}} \left( 1 + \frac{5}{4\pi} \langle\widetilde{\beta}_{2}^{2} \rangle_{A} \right)	~,
\end{equation}
where $\langle r^{2} \rangle_{A}^{\mathrm{sph}}$ is the mean-square charge radius of a spherical liquid-drop-like nucleus with mass, $A$~\cite{Otten1989}. This is calculated using the modified liquid-drop model of Ref.~\cite{Myers1983} and the updated parameter set of Ref.~\cite{Berdichevsky1985}.
Assuming that $\langle \widetilde{\beta}_{2}^{2} \rangle_{212}=0.062(5)$, from the Grodzins-Raman rule~\cite{Raman2001} and Equation~\ref{eq:beta2_laser}, and using the tabulated $\Delta\langle r^{2} \rangle_{A,212}$ values from Ref.~\cite{Otten1989}, ${\langle \widetilde{\beta}_{2}^{2} \rangle_{A}}$ can be deduced. All of the derived values for $\langle \beta^{2}_{2} \rangle^{1/2}$ and $\langle \widetilde{\beta}^{2}_{2} \rangle^{1/2}$ as a function of mass number are plotted in Figure~\ref{fig:deformation} for comparison.
A good level of consistency between the two deformation parameters is observed.
Furthermore, the values obtained from the isotope shift at the heaviest masses remain relatively constant apart from the odd-even staggering effect.
This might be considered as due to dynamical effects about a spherical shape, i.e. vibration, whereas increase of ${\langle \widetilde{\beta}_{2}^{2} \rangle}$ in the lightest isotopes points towards an onset of deformation in the ground state.

\begin{figure*}[tb]
\centering
\includegraphics[width=\textwidth]{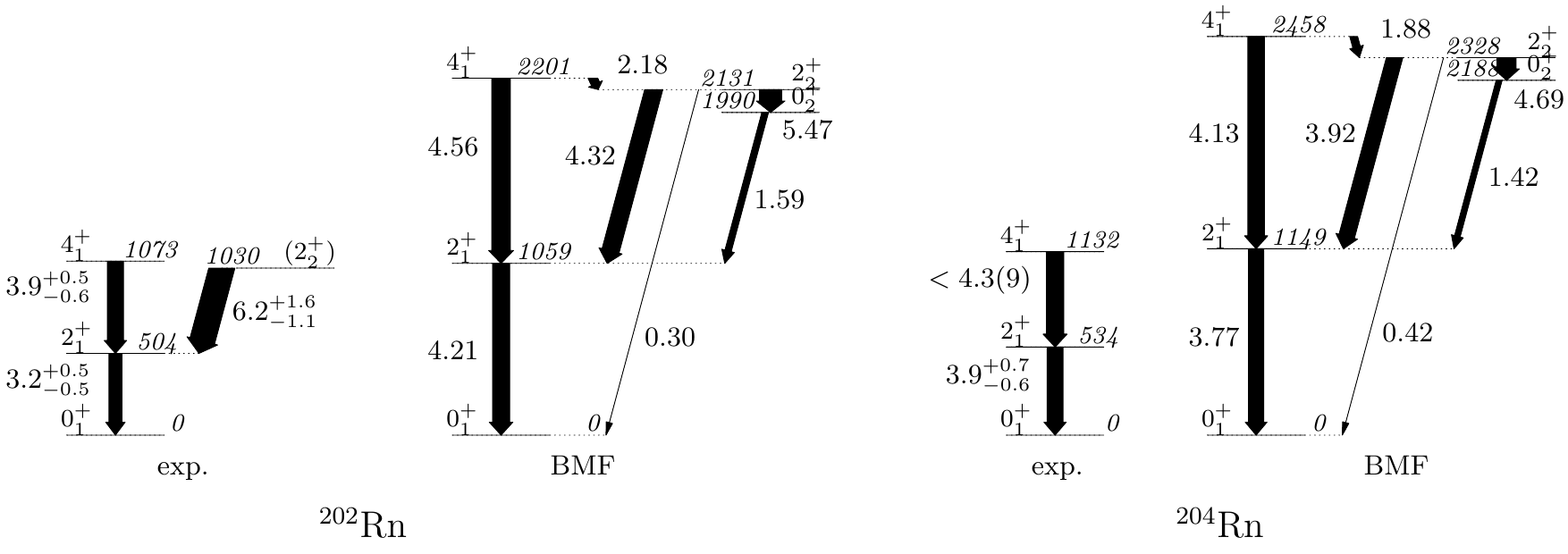}
\caption{Comparison of the results of beyond-mean-field calculations and experimental energy levels (italics; in units of keV) and transitional quadrupole moments, $|Q_{t}|$ (in units of $e$b). The width of the arrows are proportional to $|Q_{t}|$. Only states up to $4^{+}_{1}$ and $2^{+}_{2}$ are included for clarity of presentation.}
\label{fig:BMF_levels_qt}  
\end{figure*}

A less model-dependent picture of the quadrupole collectivity is the transitional quadrupole moment, $Q_{t}$, related to the experimental matrix elements by the following relationship:
\begin{equation} \label{eq:qt}
Q_{t} ( I_{i} \to I_{f} ) = \frac{ \langle I_{f} \| E2 \| I_{i} \rangle } { \langle I_{f} 0 2 0 | I_{i} 0 \rangle } \cdot \sqrt{ \frac{16\pi}{5\left(2I_{f}+1\right)} }~	,
\end{equation}
where $\langle I_{f} 0 2 0 | I_{i} 0 \rangle$ is the Clebsch-Gordan coefficient. The values deduced from the current experiment are given on the level schemes of Figure~\ref{fig:BMF_levels_qt}. Here, we can observe that, as a function of increasing spin, $Q_{t}$ remains constant in $^{202}$Rn as far as the data extends.
This can be an indicator that these states form a single rotational band, but the current level of uncertainty and number of data is not enough to make firm conclusions within such a simple picture.
The alternative and equally simplistic picture of an harmonic vibrator gives the relationship between transition strengths of two-phonon ($N_{ph}=2$) and one-phonon ($N_{ph}=1$) states as:
\begin{equation} \label{eq:phonon}
\frac{ B(E2; J^{+}_{N_{ph}=2} \to J^{+}_{N_{ph}=1}) }{ B(E2; J^{+}_{N_{ph}=1} \to J^{+}_{N_{ph}=0}) } = 2		~.
\end{equation}
While this is consistent with the ${B(E2; 4^{+}_{1} \to 2^{+}_{1})}$ from this experiment, it is at odds with the observation of a strong ${B(E2; 2^{+}_{2} \to 2^{+}_{1})}$ value, pointing to the fact that these nuclei can not be described as simple vibrators (see Table~\ref{tab:me}).
For these reasons, comparisons to state-of-the-art nuclear models are required to understand the behaviour of these nuclei.

Beyond mean-field calculations have recently been performed for a range of nuclei in this region~\cite{Yao2013}, having particular success in describing the electromagnetic matrix elements above and below $Z=82$ in the polonium~\cite{Kesteloot2015} and mercury~\cite{Bree2014,Wrzosek-Lipska2015} isotopes.
In these calculations, self-consistent mean-field wave functions are generated within the Hartree-Fock (HF) + BCS framework with a Skyrme energy-density functional.
These are then projected to particle number and angular momentum, before being mixed by the generator coordinate method (GCM) to give physical states.
The pure mean-field wave-functions are constrained to axial symmetry. 
While the parameters of the microscopic Skyrme interaction are fitted to large sets of data, the extraction of nuclear observables from the projected mean-field states can be considered parameter free.
This is very advantageous when making predictions of behaviour where experimental data is not already present.
As can be seen in Figure~\ref{fig:BMF_levels_qt}, and has also been observed in the polonium~\cite{Kesteloot2015} and mercury~\cite{Bree2014,Wrzosek-Lipska2015} isotopes, the absolute values of the energy levels predicted by the BMF model~\cite{Yao2013} appear vastly overestimated, but the general pattern is reproduced.
The prediction of a $0^{+}_{2}$ state close in energy to the $2^{+}_{2}$ state was not able to be tested in this experiment.
An observation of this state, along with its de-excitation branching ratio, would give a further test to the model.
What is interesting to note is the prediction of a very weak $2^{+}_{2} \to 0^{+}_{1}$ decay branch, consistent with the observed data, with no need to invoke arguments for a forbidden $\Delta N_{ph}=2$ transition in the harmonic-vibrator model.

\begin{figure}[tb]
\centering
\includegraphics[width=\columnwidth]{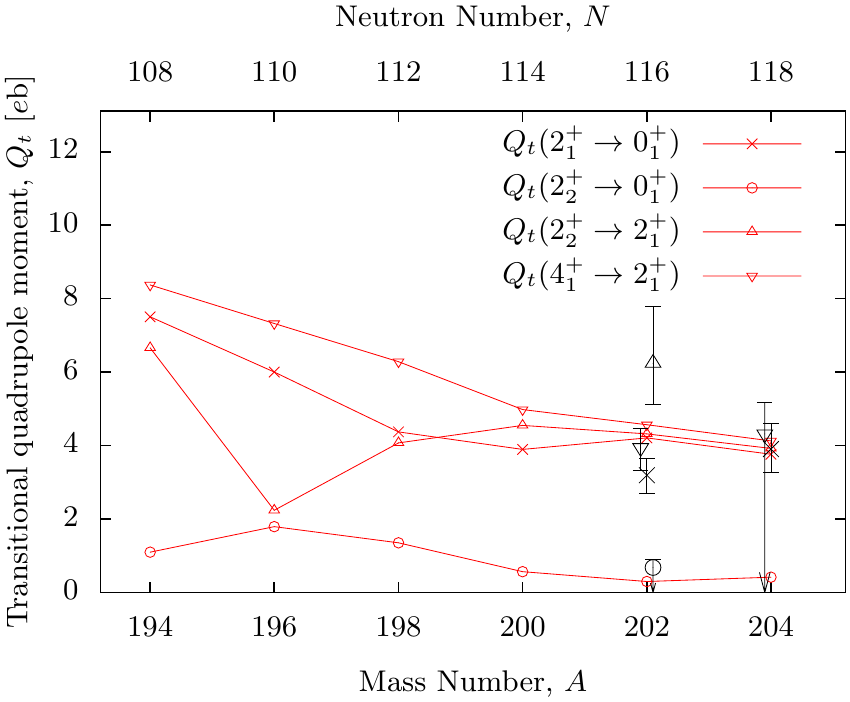}
\caption{(Color online) Experimental $Q_{t}$ values in black compared to those predicted by beyond-mean-field calculations in red, connected by lines to guide the eye. The upper limits in the data are indicated by the downward pointing arrows.}
\label{fig:BMF_Qt}  
\end{figure}

In Figure~\ref{fig:BMF_Qt}, the transitional quadrupole moments are compared for a range of nuclei extending to $A\ge194$.
An increase in collectivity for the lighter radon isotopes is predicted by an increasing $Q_{t}(2^{+}_{1} \to 0^{+}_{1})$ value, consistent with the $E(2^{+}_{1})$ systematics and isotope shift measurements~\cite{Borchers1987,Georg1995,*Georg1997}.
A more stringent test of this model would come from measurements of non-yrast and inter-band $Q_{t}$ values, which show more significant deviations when approaching mid-shell.
The current production rates at ISOLDE do not allow Coulomb-excitation experiments to be extended to isotopes lighter than $^{200}$Rn.
However, with the higher beam energies afforded by HIE-ISOLDE~\cite{Lindroos2008}, multiple-step Coulomb-excitation experiments will be able to provide a complete set of electromagnetic matrix elements for the heavier-mass isotopes, which may act as a verification of model predictions at lower masses.

\section{Conclusions}

Coulomb excitation of secondary, post-accelerated radioactive beams of $^{202}$Rn and $^{204}$Rn has been performed at the REX-ISOLDE facility in CERN.
${B(E2;2^{+}_{1} \rightarrow 0^{+}_{1})}$ values have been extracted in both $^{202}$Rn and $^{204}$Rn and limits on $Q_{s}(2^{+}_{1})$ have been determined.
In $^{202}$Rn, population of the $2^{+}_{2}$ and $4^{+}_{1}$ states was observed, allowing  the extraction of ${B(E2;4^{+}_{1} \rightarrow 2^{+}_{1})}$ and ${B(E2;2^{+}_{2} \rightarrow 2^{+}_{1})}$ values in this nucleus.
While the excitation energies of the observed states in these Rn isotopes coincide with that expected of a simple quadrupole vibrator structure, the $2^{+}_{2} \to 2^{+}_{1}$ transition strength does not support such an interpretation. 
The results have been compared to recent beyond-mean-field calculations~\cite{Yao2013}.
While the energy levels seem to be unreasonably expanded, the relative behaviour and absolute transitions strengths shows consistency between experiment and the model description.
A more sensitive test of the nuclear shape would come from the spectroscopic quadrupole moment, $Q_{s}(2^{+}_{1})$, but the precision from this experiment is not sufficient to distinguish between \mbox{oblate-,} prolate- and spherical-like charge distributions.
Extending $B(E2)$ measurements to lighter, more exotic nuclei, where shape-coexistence effects and ground-state deformations are expected to be stronger due to the parabolic behaviour of the intruding structure, would test the BMF description further.
Observation of a $0^{+}_{2}$ state is still lacking in the light radon isotopes.
New experiments at higher beam energy would increase the probability of populating this state, should it exist.
A future coupling of Miniball with the SPEDE electron detector in Coulomb-excitation experiments~\cite{Konki2013,spede} may allow direct detection of the $E0(0^{+}_{2} \to 0^{+}_{1})$ decay.
This will lead not only to a placement of the $0^{+}_{2}$ state in energy, but also to the determination of the $E2(0^{+}_{2} \to 2^{+}_{1})/E0(0^{+}_{2} \to 0^{+}_{1})$ branching ratio, key to distinguishing between an intruder and a phonon structure.
Indeed, few-nucleon transfer reactions such as ($t$,$p$) and ($d$,$p$) also have the ability to populate such excited $0^{+}$ states and could be utilised to elucidate their nucleon configuration.

\begin{acknowledgements}
We acknowledge the support of the ISOLDE Collaboration and technical teams.
This work was supported
by GOA/2010/10 (BOF KULeuven),
by the IAP Belgian Science Policy (BriX network P6/23 and P7/12),
by the U.K. Science and Technology Facilities Council (STFC),
by the German BMBF under contract Nos. 05P12PKFNE, 06DA9036I and 05P12RDCIA,
by the Academy of Finland (Contract No. 131665),
by the European Commission through the Marie Curie Actions call PIEFGA-2008-219175 (J.P.).
L.P.G. acknowledges FWO-Vlaanderen (Belgium) as an FWO Pegasus Marie Curie Fellow.
\end{acknowledgements}

\bibliographystyle{aip_doi}
\bibliography{RadonCoulex}

\end{document}